# Current-induced CrI$_3$ surface spin-flop transition probed by proximity magnetoresistance in Pt

Tang Su,[1,2] Mark Lohmann[2], Junxue Li[2], Yadong Xu[2], Ben Niu[2], Mohammed Alghamdi[2], Haidong Zhou[3], Yongtao Cui[2], Ran Cheng[4], Takashi Taniguchi[5], Kenji Watanabe[5], and Jing Shi[2]

1. International Center for Quantum Materials, School of Physics, Peking University, Beijing 100083, P. R. China
2. Department of Physics and Astronomy, University of California, Riverside, CA 92521, USA
3. Department of Physics and Astronomy, University of Tennessee, TN 37996, USA
4. Department of Electrical and Computer Engineering, University of California, Riverside, CA 92521, USA
5. National Institute for Materials Science, 1-1 Namiki, Tsukuba, 305-0044, Japan

**Email:** jing.shi@ucr.edu

By exploiting proximity coupling, we probe the spin state of the surface layers of CrI$_3$, a van der Waals magnetic semiconductor, by measuring the induced magnetoresistance (MR) of Pt in Pt/CrI$_3$ nano-devices. We fabricate the devices with clean and stable interfaces by placing freshly exfoliated CrI$_3$ flake atop pre-patterned thin Pt strip and encapsulating the Pt/CrI$_3$ heterostructure with hexagonal boron nitride (hBN) in a protected environment. In devices consisting of a wide range of CrI$_3$ thicknesses (30 to 150 nm), we observe that an abrupt upward jump in Pt MR emerge at a 2 T magnetic field applied perpendicularly to the layers when the current density exceeds $2.5 \times 10^{10}$ A/m$^2$, followed by a gradual decrease over a range of 5 T. These distinct MR features suggest a spin-flop transition which reveals strong antiferromagnetic interlayer coupling in the surface layers of CrI$_3$. We study the current dependence by holding the Pt/CrI$_3$ sample at approximately the same temperature to exclude the joule heating effect, and find that the MR jump increases with the current density, indicating a spin current origin. This spin current effect provides a new route to control spin configurations in insulating antiferromagnets, which is potentially useful for spintronic applications.

**Introduction**

After the discovery of single layer graphene[1,2], the research on van der Waals (vdW) materials and heterostructures including recently discovered vdW magnetic materials has remained extremely active [3–7]. $CrI_3$, the most studied vdW chromium trihalide, has been at the center of the spotlight due to its unique tunable magnetic and magneto-electric properties. In its bulk crystal form, $CrI_3$ shows a ferromagnetic behavior, with a band gap of 1.2 eV and a Curie temperature of 61 K[8–10]. The magnetic anisotropy is perpendicular to the two-dimensional (2D) atomic layers. Similar to other chromium trihalides, the super-exchange interaction mediating the magnetic ions separated by non-magnetic ligands is believed to be responsible for the fascinating magnetic properties[9].

In thin $CrI_3$ devices with the thickness ranging from monolayer to 20 layers, however, the magnetic properties deviate from the bulk ferromagnetic behavior. It was found that the interlayer coupling turns into antiferromagnetic (AFM) as has been probed by magneto-optic Kerr effect[6], reflectance magnetic circular dichroism[11], and tunneling magneto-resistance measurements[6],[11–13]. Particularly, in a 10-layer-thick $CrI_3$ device, the tunneling magnetoresistance between parallel and anti-parallel interlayer spin alignments can be as large as one million percent[14]. In addition, it has been shown that the interlayer coupling in ultra-thin $CrI_3$ can be modulated by applying electric field, electrostatic doping[15–17], or hydrostatic pressure[18–20]. These extraordinary behaviors manifested in thin $CrI_3$ devices suggest that the surface plays an important role.

Besides the surface layers, interfaces in heterostructures can greatly affect the properties of both $CrI_3$ and its neighboring material. Compared with conventional three-dimensional (3D) magnetic materials, 2D layered magnetic materials such as $CrI_3$ allow for easier fabrication of heterostructures with other 2D non-magnetic materials for studying various proximity effects. For example, a variety of emergent phenomena including the Rashba spin-orbit coupling (SOC)[21,22] can be induced by interfacial coupling with other layered materials. By proximity coupling $CrI_3$ with $Bi_2Se_3$ topological insulator, another vdW material, a sizable spin splitting at the Dirac point of the topological surface states was predicted, which is promising for the realization of the quantum anomalous Hall effect[23]. Graphene/$CrI_3$ is another heterostructure proposed to realize a Chern insulator[24]. Experimentally, heterostructures of ultrathin $CrI_3$ and a monolayer of $WSe_2$ have been investigated by photoluminescence[25] to show a large exchange field of 13 T. In this work, we fabricate heterostructures consisting of $CrI_3$ and Pt in which the magnetoresistance (MR) of the Pt layer serves as a detector to sense the spin state of the $CrI_3$ surface layers. We have also observed an effect of current in Pt on spin configurations in $CrI_3$ which indicates the effect of spin current generated by Pt.

**Device fabrication and transport measurements**

Previous studies[11–14] incorporated ultrathin $CrI_3$ flakes (1-20 layers) as the barrier layer in tunnel devices. The tunneling MR jumps revealed transitions among different spin configurations between atomic layers. Here, we adopt a different approach that is routinely used to study 3D magnetic insulators such as YIG[26,27] and was previously applied to study another 2D semiconducting vdW ferromagnet $Cr_2Ge_2Te_6$ (CGT)[28], i.e., leveraging proximity coupling with a thin layer of Pt. Similar proximity magneto-transport studies have also been reported in other 2D systems [25,29,30]. However, due to fabrication challenges associated with the ultra-sensitivity of $CrI_3$, heterostructure devices such as Pt on $CrI_3$ denoted as $CrI_3$/Pt have not yet been successfully fabricated and investigated. To circumvent the challenges, in this work, we first pattern a 5 nm thick Pt Hall bar by electron-beam lithography and magnetron sputtering deposition with the channel length of 6 μm and width of 4 μm on an exfoliated flake of hexagonal BN (hBN) placed on $Si/SiO_2$ substrate. The Pt Hall bar is then connected to outer Au contact pads for wire bonding. The crystals of $CrI_3$ are prepared by the chemical vapor transport method. The mixture of chromium powder and iodine pieces with appropriate ratio is sealed in an evacuated quartz tube which is placed in a tube furnace with the hot and cold ends anchored at 650 °C and 550 °C, respectively. After two days, the furnace is cooled down to room temperature and crystals are obtained on the cold end. To avoid exposing $CrI_3$ to air or moisture, we perform exfoliation and transfer in a protective environment - inside a glovebox with $O_2$ and $H_2O$ concentration < 0.1 ppm. We use the standard dry transfer method to pick up and place the chosen $CrI_3$ flake on top of the Pt Hall bar (See supplementary Fig. S1 for atomic force microscopy imaging data). The $Pt/CrI_3$ heterostructure is then capped with a larger hBN flake so that it is well encapsulated between two hBN sheets even with a 30 to 150 nm thick $CrI_3$ and a 5 nm thick Pt because of the flexibility of hBN. The device fabrication is completed inside the glovebox before it is taken out for transport measurements.

Figure 1 (a) shows the side view of the monoclinic (C2/m) crystal structure of $CrI_3$, which is the stable phase at room temperature. Figure 1 (b) is the Raman spectrum of an exfoliated $CrI_3$ flake encapsulated by hBN layers. The room temperature spectrum contains the characteristic peaks associated with the monoclinic phase, confirming the stability of hBN-encapsulated $CrI_3$ flake[31,32](see supplementary Fig. S2 for more Raman details). Figure 1(c) shows device images with (upper) and without (lower) $CrI_3$ on top of a Pt Hall bar. Figures 1(d) is the schematic cross-sectional view of an encapsulated device, and Figure 1(e) an illustration of device structure and measurement geometry. We perform DC magneto-transport measurements with a current of 2 mA (unless otherwise specified) in a Physical Property Measurement System (Dynacool, Quantum Design), with the maximum magnetic field of 14 T and the base temperature of 2 K.

**Magnetoresistance**

Figure 2(a) shows the four-terminal MR data of device **D#1** ($CrI_3$ thickness ~80 nm) with the magnetic field applied perpendicularly to the atomic layers. Superimposed on top of the overall smooth positive MR background, there are two sharp jumps located at about ±2 T that are nearly symmetric in magnetic field and have little hysteresis between the up- and down-sweeps. As will be discussed in more detail later, the MR jumps are absent if the current density is below a certain value, indicating the nonequilibrium nature of the jumps. To further understand the MR behavior of this $Pt/CrI_3$ heterostructure device, we also measure the MR of a pre-patterned Pt Hall bar fabricated on hBN without $CrI_3$ on top. The Pt-only device shows a parabolic MR as a function of the magnetic field, which is expected for non-magnetic materials originating from the Lorentz force. Due to the small differences in dimensions between different sputtered Pt Hall bars, the absolute value of the Pt resistance can vary by a small amount. Therefore, we multiply the MR data of the Pt-only device by a constant (orange solid curve) to align with the MR data of $Pt/CrI_3$ at 14 T (black solid curve) and place both curves in the same plot. At high fields (from 7 to 14 T), the MR data overlap with each other very well, and difference becomes visible only below ~ 7.5 T. We also notice that the MR jumps in $Pt/CrI_3$ occur at approximately the same fields as the highest-field jumps in the tunneling MR data reported in tunnel junction devices with 3-20 layer thick $CrI_3$[11–14]. By comparing the MR data in Figure 2(a), we conclude that the MR jumps in $Pt/CrI_3$ are caused by the presence of $CrI_3$. Since no heat treatment is performed after the $CrI_3$ flake is transferred onto Pt at room temperature, it is unlikely that diffusion of Pt atoms occurs across the interface to make CrI$_3$ conducting and produce its own MR. We also notice that after the CrI$_3$ flake in Pt/CrI$_3$ is totally degraded, the MR jumps disappear and the curve becomes parabolic just as that of the Pt-only device. Therefore, the MR jumps must stem from the conduction electrons in Pt that are sensitive to the direction of spins at the CrI$_3$-Pt interface (see supplementary Figs. S3 & S4 for images and MR curves of a few other devices).

Similar induced effects in 3D magnetic insulator/Pt heterostructures such as YIG/Pt have been previously studied[27,33]. While a complete consensus is still lacking, the anisotropic MR and spin Hall MR are two possible underlying mechanisms. Our MR data with the field rotating in the zx- and zy planes favor the former (see supplementary Fig. S5). Regardless of actual mechanism, the MR conforms with $\pm\sin^2\theta$, where $\theta$ is the angle between the magnetization in the magnetic layer and the current direction. Therefore, if a simple spin flip occurs, i.e., from $\theta$ to $\theta$ +180°, the MR would not register any change. This is also true for the spin-dependent scattering mechanism by localized moments. Therefore, the observed abrupt MR jumps qualitatively exclude this spin flip scenario, which has been identified as the mechanism of similar jumps in the magneto-optic Kerr effect[6] and tunneling MR data[11–14]. In fact, after the MR jump, there is a continuous change in our MR data as the field increases further until it finally merges to the parabolic background. To highlight the induced MR effect due to CrI$_3$, we subtract the parabolic background from the MR curve of Pt/CrI$_3$ and show the difference in Figure 2(b) (only the positive field side due to symmetry in field). It becomes more evident that the 2 T abrupt upward jump is followed by a smooth decrease before reaching saturation. Due to the short-range nature of the proximity coupling, these MR features suggest that the surface spin orientation of CrI$_3$ undergoes an abrupt change at 2 T followed by a continuous rotation towards the magnetic field. At around 8 T, the MR reaches the zero-field value, indicating full alignment of all spins towards the magnetic field. We observe no clear jumps in the Hall channel. The maximum anomalous Hall resistance signal is no larger than $10^{-6}$ of the longitudinal resistance, which may be due to the nature of the special device fabrication method adopted for the extremely sensitive CrI$_3$.

**Spin-flop transition**

Clearly, the MR jump in Pt reveals an abrupt reorientation of CrI$_3$ surface spins as the magnetic field exceeds a critical value regardless of the physical origin of the MR response. Considering the subsequent gradual change in MR, the spin orientation can only transition from the initial angle $\theta = 0°$ to an intermediate angle between 0° and 90° at the critical field (see supplementary Fig.S10 for the schematic of spin-flop transition in this heterostructure system). This behavior strongly suggests the spin-flop transition which is a characteristic of uniaxial AFM arising from the competition among the AFM exchange interaction, magnetic anisotropy, and Zeeman interaction. Depending on the relative strength of exchange interaction and magnetic anisotropy, however, the spin-flop configuration is not always energetically favorable. This is because the spin-flop critical field is $H_c = \sqrt{H_A(2H_E - H_A)}$ and the field required to fully align the spins is $H_{\text{sat}} = 2H_E - H_A$ with $H_E$ and $H_A$ being the exchange and the anisotropy fields, respectively. If $H_E < H_A$, then $H_{\text{sat}} < H_c$, a direct transition to the fully aligned phase via spin-flip is preferred, which had been observed in previous works of CrI$_3$. An intermediate spin-flop phase between the collinear AFM and spin aligned phases is energetically possible[34] only when $H_c < H_{\text{sat}}$, which requires $H_E > H_A$. Therefore, the following two properties can be inferred from the observation of the current-induced spin-flop transition. First, there is strong interlayer AFM coupling in the surface layers of CrI$_3$. Second, the AFM coupling and/or anisotropy depend on current so that the condition $H_E > H_A$ can be met by applying a large current.

To further confirm the spin-flop nature of the MR jump, we compare the MR behavior with the external magnetic field $H$ oriented both along the out-of-plane and in-plane directions, i.e., $\theta_H$ at 0° and 90°, respectively (Figure 2(c)). The abrupt MR jump is absent for $\theta_H$=90°. Moreover, by varying $\theta_H$ from 0° to 90°, we show the angular dependence of MR in Figure 2(d). Below a certain angle ($\theta_H \sim 60°$), the MR jump is still present at approximately the same critical field of 2 T. Above 60°, the MR response rises smoothly and then approaches saturation. The saturation field is approximately the same as that appeared in the tunneling MR when the field is directed in-plane[11]. The absence of abrupt MR jump above a critical $\theta_H$ is another characteristic feature of the AFM spin-flop transition which is determined by the ratio of the anisotropy over the exchange interaction[35].

We have also measured the MR evolution as a function of the system temperature. As shown in Figure 3(a), the MR jump magnitude decreases with an increasing temperature and completely disappears above 45 K. Even if we take into account the sample temperature rise due to joule heating (temperature calibration and more detailed temperature dependence are shown in supplementary Figs. S6 & S7) which is less than 5 K calibrated by the resistive thermometry of the Pt underneath CrI$_3$ at ~ 45 K, the actual sample temperature (45-50 K) is still significantly lower than the bulk CrI$_3$ $T_C$ of ~ 61 K. This suggests a lower magnetic ordering temperature of the surface layers, which is consistent with the recent finding of a lower surface ordering temperature studied by magnetic force microscopy [36]. As the temperature increases, the spin-flop field $H_c$ decreases since both $H_A$ and $H_E$ vanish at $T_C$, so does $H_c = \sqrt{H_A(2H_E - H_A)}$. In the meantime, as the magnitude of the Néel vector becomes smaller at higher temperatures, the magnitude of the MR jump also decreases. These features are consistent with the observations in Figs. 3(a) and (b).

The same abrupt spin-flop features at 2 T are observed in all 10 Pt/CrI$_3$ devices with the CrI$_3$ thickness ranging from 30 nm to 150 nm, which justifies the reproducibility of the effect (see supplementary Figs. S3 & S4). In relatively thick devices, the interior layers adopt the rhombohedral phase which favors ferromagnetic interlayer coupling [36]. However, since the Pt MR is only sensitive to the surface spins of CrI$_3$ and we always observe the MR jump at 2 T in these devices, we conclude that the surface layers in all devices show strong AFM interlayer coupling regardless of the device thickness.

**Nonequilibrium origin: current dependence**

As briefly mentioned earlier, the MR jump in Pt/CrI$_3$ devices only occurs when the current density exceeds a certain value. To study the current effect, we have carried out a series of current dependence measurements with DC currents $I$ ranging from 10 µA to 4.5 mA, which corresponds to $5\times10^8 - 2.25\times10^{11}$ A/m$^2$ in current density, spanning more than two orders of magnitude. In this DC current range, we observe no irreversible resistance or magnetoresistance changes, indicating little damages or current annealing effect. Figure 4(a) shows the MR results measured at the same system temperature of 2 K. For $I < 0.5$ mA, no MR jump can be identified (not shown). When $I >$ 0.5 mA, the MR jump emerges; it reaches the maximum at ~2 mA and then starts to broaden and decrease at larger currents. The initial $I$-dependence trend strongly suggests that the spin-flop transition is an induced phenomenon which requires a sufficiently large current. On the other hand, the decreasing trend of the MR jump is indicative of sample heating at large currents. As $I$ increases, it produces joule heating to raise the sample temperature above the system set temperature. In comparison with Figure 3(a), as the current exceeds 2 mA, the higher sample temperature first suppresses and eventually quenches the MR jump when it reaches the magnetic ordering temperature of CrI$_3$. If the MR responses for different DC currents are directly compared for the same system temperature, these two effects are entangled (See supplementary Figs. S7 and S8), which makes the overall *I*-dependence difficult to analyze.

We use the Pt four-terminal resistance in Pt/CrI$_3$ device as a thermometer to accurately determine the temperature of the Pt layer. The sample temperature profile under large currents clearly depends on some details such as the sample geometry and thermal conductivity of the relevant materials. However, the determined Pt temperature should be a good measure of the CrI$_3$ surface temperature. To separate the joule heating effect under different currents, for each current, we adjust the system temperature to stabilize the sample temperature at the same target value monitored by the Pt resistance thermometry(see details in Fig. S6). The higher the current is, the lower the system temperature must be set. In Figure 4(b), we plot several representative MR curves for different DC currents up to 2 mA at the calibrated sample temperature of ~ 25 K. To keep the sample temperature at 25 K, we are limited to the largest current of 2 mA which requires the system temperature to be set at 2 K, the minimum system temperature. The current dependence of the spin-flop MR jump magnitude is shown in Figure 4(c). It is clear that at both 2 K and 25 K, no MR jump is observed if the current is below 0.5 mA. Therefore, unlike in conventional AFM materials, the spin-flop transition observed here is not an equilibrium phenomenon, but requires large currents. The larger MR jump magnitude under larger currents is attributed to the larger deviation angles of the spin orientations from the easy axis right after the spin-flop transition, suggesting a spin current (to be discussed in the next section) rather than a heating effect. At the sample temperature of 25 K, the MR jump has a monotonic current dependence up to the maximum current of 2 mA.

**Spin current effect**

Next let us examine three possible mechanisms responsible for this current-induced spin-flop phenomenon. First, a trivial effect of the electric current is the generation of a magnetic field or Oersted field. However, the oersted field is directed parallel to the surface of CrI$_3$; therefore, it does not cause spins to flop. As shown in Figure 2(d), spin-flop requires the field to be applied along the easy-axis direction, i.e., perpendicular to the surface of CrI$_3$. In addition, a 2 mA current only produces a ~ 1 mT field, far smaller than the spin-flop field. Hence, the oersted field is irrelevant to the MR jump. Second, a charge current in Pt and other conductors with strong SOC produces a spin current with spin polarization $\sigma$ parallel to the atomic layers via the spin Hall effect, which carries a flow of angular momentum. When a spin current enters a ferromagnet, it is known to exert spin-orbit torques, e.g., field-like and damping-like torques, to cause interesting magnetization dynamics such as rotation and switching of the magnetization, which has been vigorously studied recently[37,38]. Although the surface layers of $CrI_3$ are antiferromagnetically coupled, the same torques can act on the surface sublattice spins. Similar to the oersted field, however, the equivalent field of the field-like torque, $\boldsymbol{H_{FL}} \sim \boldsymbol{\sigma}$, is parallel to the atomic layers of $CrI_3$ and acts like an effective field along y-axis, therefore does not cause spins to flop either. As shown in Fig. S11, the effective field, $\boldsymbol{H_{DL}} \sim \boldsymbol{m} \times \boldsymbol{\sigma}$, of the damping-like torque, on the other hand, can cause spins to rotate away from the easy-axis orientations and flop to an intermediate angle if the torque is sufficiently strong (more evidence is shown in supplementary Fig. S12). Interestingly, the minimum current of 0.5 mA for spin-flop corresponds to $2.5 \times 10^{10}$ A/m$^2$ in current density, which is on the same order of magnitude in ferromagnetic devices showing spin-orbit torque-induced switching of magnetization[39,40]. Recently, a mechanism taking into consideration of all allowed spin-orbit torques has been proposed to produce a reduction in magnetic anisotropy in $Fe_3GeTe_2$ (FGT) [41], a metallic vdW ferromagnet. In $CrI_3$, the consequence of spin-orbit torques may be equivalent to a similar anisotropy reduction and therefore $H_A$ reduction. If the current reaches a critical value, $H_E > H_A$ is fulfilled, and the spin-flop is resulted. One difference in $CrI_3$ is that it has AFM interlayer coupling. Another major difference is that FGT itself is a spin current source, while spin current needs to be externally injected to $CrI_3$ just as in other insulating ferromagnets[39]. Third, for large DC currents, joule heating produces a vertical temperature gradient, consequently, a spin current associated with a downward magnon flow. Even if we maintain the same surface temperature of $CrI_3$ under different currents, large currents produce greater temperature gradients, which results in larger accumulation of magnons at the bottom surface. This effectively reduces magnetic anisotropy in $CrI_3$ which can cause spins to flop. The last two mechanisms may be both involved in the spin-flop.

In summary, we have observed a current-induced sharp jump followed by a smooth decrease in the proximity MR of Pt in Pt/$CrI_3$ heterostructures. The MR jump is attributed to a spin-flop transition of the AFM coupled $CrI_3$ surface layers. The AFM interlayer coupling is found to exist in the surface layers of $CrI_3$ devices over a certain range of thicknesses. By carefully excluding the joule heating effect, we conclude that the spin-flop transition is caused by spin currents injected from Pt via the spin Hall effect and driven by the vertical temperature gradient in $CrI_3$. Our work demonstrates a new spin current induced effect in layered magnetic heterostructures which is potentially useful for 2D spintronic applications.

## Acknowledgements

We thank Wei Han, Mohammed Aldosary, Wei Yuan, and Victor H. Ortiz for useful discussions. Low-temperature transport measurements and data analyses were supported by DOE BES Award No. DE-FG02-07ER46351. Construction of the pickup-transfer optical microscope, device nanofabrication, and device characterization were supported by NSF-ECCS under Awards No. 1202559 NSF-ECCS and No. 1610447. H.D.Z. thanks the support from NSF-DMR with grant number NSF-DMR-1350002. K.W. and T.T. acknowledge support from the Elemental Strategy Initiative conducted by the MEXT, Japan and the CREST (JPMJCR15F3), JST.

## Author contributions

JS supervised the project. TS performed device fabrication, characterization, and transport measurements. ML, YDX, and MA participated in the experiments. HDZ provided $CrI_3$ crystals. BN and YTC performed low-temperature magnetic force microscopy on $CrI_3$ flakes. TT and KW provided hBN crystals. RC and JXL helped data analysis and provided theoretical guidance.

**Figure captions**

**Figure 1. CrI$_3$ structure and device schematics.** (a) Side view of CrI$_3$ layers in the monoclinic (C2/m) crystal structure. (b) Room-temperature Raman spectrum of exfoliated CrI$_3$. The presence of the characteristic peaks from different vibration modes (labeled) indicates good quality of exfoliated CrI$_3$. (c) Patterned Pt devices with (upper) and without (lower) CrI$_3$ flake on top. (d) Schematic cross-sectional view of Pt/CrI$_3$ device. (e) Device geometry for transport measurements.

**Figure 2. MR results in Pt/CrI$_3$ at 2 K**. (a) Four-terminal MR of Pt in devices with (black solid curve) and without (orange solid curve) CrI$_3$ on top with magnetic field perpendicular to the device. (b) Difference between the two MR curves in (a) for positive fields. MR difference data over the full field range is shown in the inset. Blue and red arrows illustrate the sublattice spin configurations in different regions, i.e., left: AFM; middle: spin-flopped; right: aligned. (c) MR curves with the magnetic field $H$ directed out-of-plane ($\theta_H=0°$) and in-plane ($\theta_H=90°$). (d) Evolution of MR curves (positive fields) with external magnetic field direction varying from out-of-plane to in-plane. In (c) and (d), the curves are shifted vertically for clarity and the scale bars represent an MR ratio of 5x10$^{-5}$.

**Figure 3. Temperature dependence of MR signal**. (a) Selected MR curves from 2 K to 45 K after the parabolic MR background is subtracted. (b) Magnitude of the MR jump (blue squares), $\Delta R_{SF}$, and the critical field, $\mu_0H_c$, at the jump (red circles) as a function of system temperature. The MR jump disappears above the system temperature of 45 K.

**Figure 4. Current dependence of MR in D#2**. (a) MR curves taken with different DC currents at the system temperature of 2 K. (b) MR curves with different DC currents $I$. To exclude heating effect, each curve is taken at the system temperature which gives the same $R_{xx}(H=0)$ reading to ensure the same actual sample temperature. The MR curves measured with $I$ =2 mA, 1.5 mA, 1 mA, 0.75 mA, and 0.5 mA are taken at the system temperatures of 2, 15, 18, 21, and 23 K,

respectively, which corresponds to the same actual sample temperature of ~ 25 K. (c) Magnitude of the MR jump $\Delta R_{SF}$ read from data in (b) as a function of *I*.

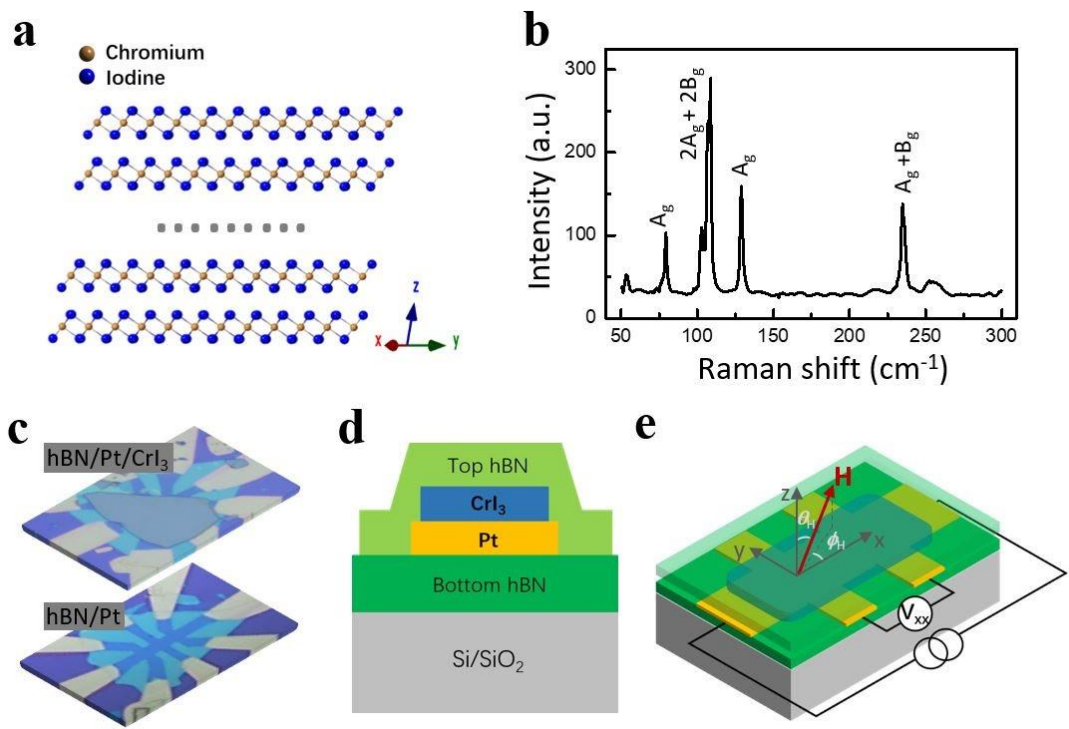

**Figure 1**

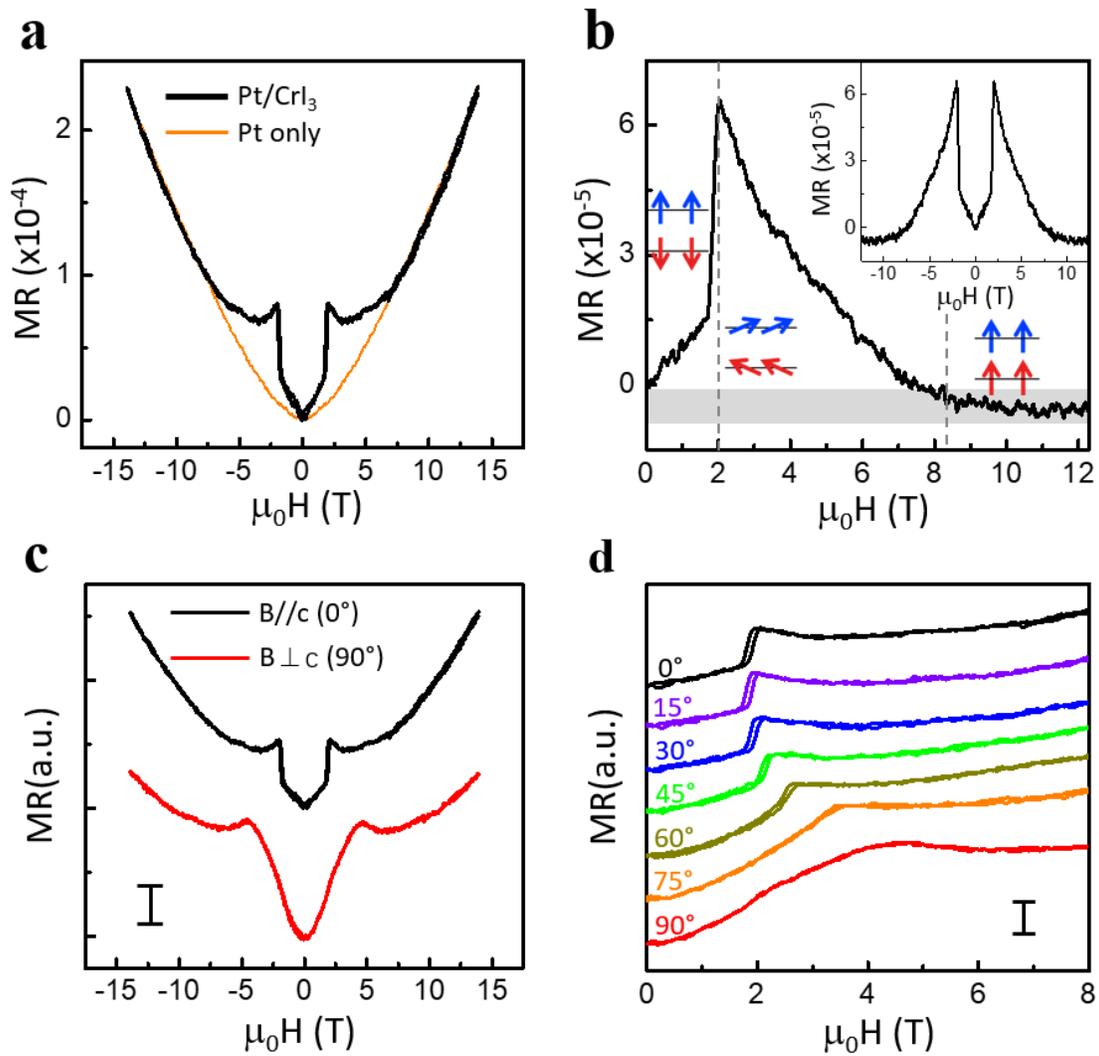

**Figure 2**

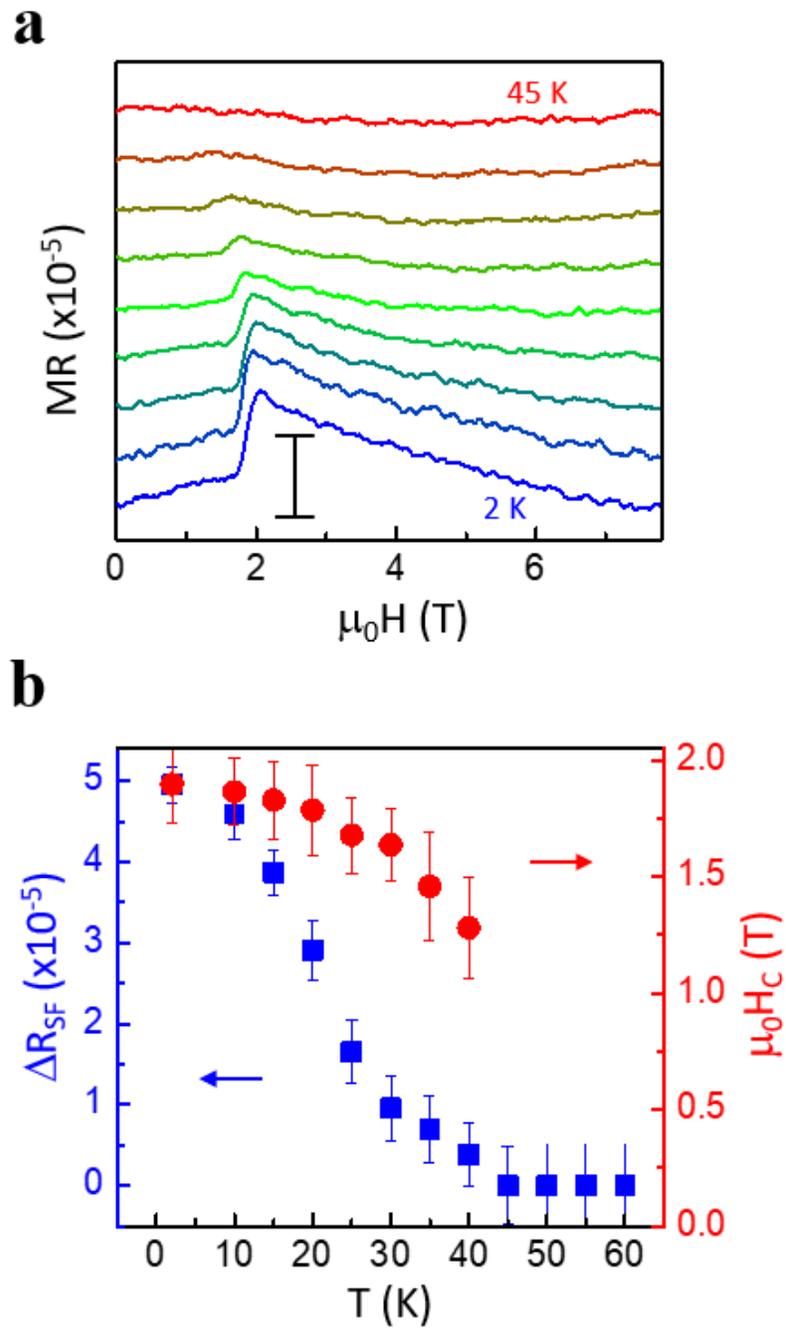

**Figure 3**

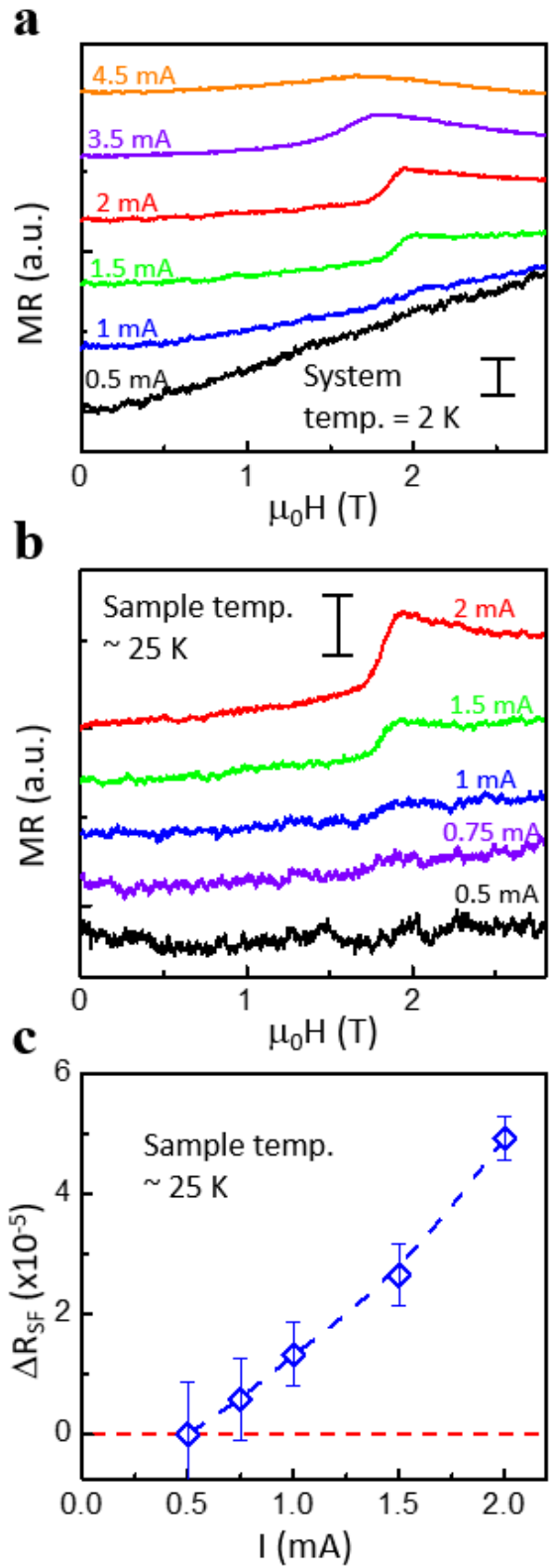

**Figure 4**